\documentclass[aps,prd,amsmath,amssymb,preprintnumbers,onecolumn,11pt,nofootinbib]{revtex4}
\usepackage{mathtools} 
\usepackage[a4paper,left=20mm,right=20mm,top=25mm,bottom=25mm]{geometry}
\usepackage{mathrsfs}
\usepackage{graphicx,epsfig}
\usepackage{color}
\usepackage{fancyhdr}
\usepackage{multirow}
\usepackage{tikz}
\usepackage{feynmp}
\usepackage{mathpazo}
\usepackage{bbold}
\usepackage{natbib}
\usepackage{diagbox} 

\usepackage{amsfonts}
\usepackage{bm}
\usepackage{xcolor}

\definecolor{CP3}{cmyk}{0,0.88,0.77,0.40}

\newcommand\be{\begin{equation}}
\newcommand\ee{\end{equation}}
\newcommand\ba{\begin{eqnarray}}
\newcommand\ea{\end{eqnarray}}

\newcommand\e{{\rm e}}
\newcommand\de{\delta}

\newcommand\al{\alpha}
\newcommand\bt{\beta}
\newcommand\pa{\partial}

\newcommand{\z}{{\cal Z}}

\renewcommand{\(}{\left(}
\renewcommand{\)}{\right)}
\renewcommand{\[}{\left[}
\renewcommand{\]}{\right]}

\newcommand{\lm}{\lambda}
\newcommand{\sig}{\sigma}

\newcommand{\m}{{\cal M}}

\newcommand\Cp{C_{,\phi}}
\newcommand\CX{C_{,X}}

\newcommand\tCp{\Gamma}
\newcommand\tCpp{\Gamma_{, \phi}}
\newcommand\tCpX{\Gamma_{, X}}
\newcommand\tCX{\Xi}
\newcommand\tCXp{\Xi_{, \phi}}
\newcommand\tCXpp{\Xi_{, \phi\phi}}
\newcommand\tCXpX{\Xi_{, \phi X}}
\newcommand\tCXX{\Xi_{, X}}
\newcommand\tCXXX{\Xi_{, X X}}
\newcommand\tQ{\tilde{Q}_0}
\newcommand\tDQ{\delta\tilde{Q}}

\newcommand\fy{{(\phi)}}
\newcommand{\dm}{{c}}
\begin{document}

\title{Coupled dark energy model inspired from general conformal transformation}
\author{Wittaya Thipaksorn $^{1}$}
\email{wittayat58@nu.ac.th}

\author{Stharporn Sapa$^{2, 3}$}
\email{stharporn@northern.ac.th}

\author{Khamphee Karwan$^{1,3}$}
\email{khampheek@nu.ac.th}

\affiliation{\footnotesize $^{1}${The Institute for Fundamental Study \lq\lq The Tah Poe Academia Institute\rq\rq, \\
Naresuan University, Phitsanulok 65000, Thailand}}

\affiliation{\footnotesize $^{2}${Northern College, Tak 63000, Thailand}}

\affiliation{\footnotesize $^{3}${Thailand Center of Excellence in Physics, Ministry of Higher Education, Science, \\
Research and Innovation, Bangkok 10400, Thailand}}


\begin{abstract}
We study the coupled dark energy model constructed from the general conformal transformation in which the coefficient of the conformal transformation depends on both the scalar field and its kinetic term.
Under this conformal transformation, the action for subclass of degenerate higher-order scalar-tensor (DHOST) theories is related to the Einstein-Hilbert action.
The evolution of the background universe has the scaling fixed point which corresponds to acceleration of the universe at late time.
For the choices of parameters which make the late-time scaling point stable, the fixed point corresponding to $\phi$-matter-dominated-era ($\phi$MDE) is a saddle point,
and the universe can evolve from radiation dominated epoch through $\phi$MDE before reaching the scaling point at late time with the cosmological parameters which satisfy the observational bound.
During the $\phi$MDE, the effective equation of state parameter is slightly positive,
so that one of possible mechanisms for alleviating the $H_0$ tension can be achieved.
In this coupled dark energy model, the effective gravitational coupling for dark matter perturbations on small scales can be smaller than that  in the $\Lambda$CDM model.
Therefore a growth rate of the dark matter perturbations is suppressed compared with the $\Lambda$CDM model,
which implies that the $\sigma_8$ tension could be alleviated. 

\end{abstract}

\maketitle

\section{Introduction}
\label{sec1}

One of challenges in cosmology is to explain the cosmic acceleration at late time which could be a consequence of unknown forms of dark energy or deviation from Einstein gravity on large scales \cite{Copeland:2006,Clifton:2011,Bamba:2012,Huterer:2017,Ishak2018,Zhang2021}.
To avoid the cosmological constant problem due to extremely large difference of the value of cosmological constant from theoretical predictions and from observations,
dark energy should be dynamical energy component.
However, the dynamical dark energy encounters the problem why the energy density of dark energy is comparable with that for dark matter at late time even though they evolve differently.
This problem of dark energy is the coincidence problem which could be alleviated if there are attractors corresponding to the cosmic acceleration at late time \cite{zlatev1,zlatev2}.
Such attractors can exist under the assumption that dark energy is coupled to cold dark matter (CDM) \cite{Amendola:2000,Wang:2016,ElsaM:1903,Patrocinio:2021}.

Possible models of coupled dark energy are inspired from the frames transformation in theories of gravity.
The interaction between dark energy and CDM can be inspired from the conformal transformation \cite{Faraoni:99,Kaiser2010,Wright:2016,Teixeira:2019,Gal'tsov:20} and disformal transformation \cite{Bettoni:2013,Sakstein:14,Shinji:2015,Sakstein:15,vandeBruck:15,Karwan:2017,Dusoye:2021,Gomez:2021jbo,Gomez:2022okq}.
When the transformation coefficient depends on the scalar field only,
the coupling due to conformal transformation leads to a coupling between field and energy density of CDM which corresponds to a energy transfer between the dark components \cite{Amendola:2000,vandeBruck:2016}.
For the coupling from disformal transformation, the velocity perturbations and the time derivative of density perturbation in CDM can be eliminated from the effective coupling term in the evolution equation on small scales.
As a result, the effective coupling term depends only on the perturbations in energy density of CDM similar to the coupling from the conformal transformation \cite{skd:18}.
The effects of the disformal coupling appear at the background level through the modification of the cosmic expansion and the coefficient of the perturbed coupling term. 
The growths of the matter perturbations on small scales for these two types of coupled dark energy models are higher than that for $\Lambda$CDM model \cite{Pettorino:2008,skd:18}.

The interaction between the dark components can also be arisen from a coupling betweenthe a field derivative $\partial_\mu\phi$ and a CDM four velocity $u^\mu$ in the action \cite{Pourtsidou:2013nha,Boehmer:2015kta,Boehmer:2015sha,Skordis:2015yra,Koivisto:2015qua,Pourtsidou:2016ico,Dutta:2017kch,Linton:2018,Kase:2019veo,Chamings:2019kcl,Amendola:2020}.
This form of the coupling can lead to a pure momentum transfer between the dark components. 
Interestingly, it can reduce the effective gravitational coupling relevant to the growth of CDM perturbations,
and therefore a growth of matter perturbations on small scales is suppressed \cite{Pourtsidou:2013nha,Pourtsidou:2016ico,Kase:2019mox,Amendola:2020}.

The suppression of the matter growth rate in this class of coupled dark energy models potentially alleviates $\sigma_8$ tension,
where $\sigma_8$ is the amplitude of matter perturbation inside comoving radius $8 h^{-1}$Mpc, $h = H_{0}/100$ and $H_0$ is the Hubble expansion rate at present.
Even though the $\Lambda$CDM model  satisfies the observational data well,
it suffers from the tension of the $\sigma_8$ between the Cosmic Microwave Background (CMB) and shear lensing analyses \cite{Heymans:2012,Hildebrandt:2017}.
The estimated $\sigma_8$ from CMB data is larger than that from shear lensing analyses.
Since the $\sigma_8$ tension arises in $\Lambda$CDM model,
the $\sigma_8$ tension could be solved if the growth of matter perturbations is suppressed compared with that for $\Lambda$CDM.

The other interesting feature of the coupled dark energy models is the existence of the scaling fixed points \cite{Amendola:2006,Amendola:2020}.
These points can describe the accelerated expansion of the universe at late time as well as the matter dominated epoch.
The scaling point that can represent the matter dominated epoch is the $\phi$MDE point in which there is a small density fraction of dark energy during matter domination.
The coincidence problem could be alleviated if the universe can evolve from the radiation dominated epoch through the $\phi$MDE which should be a saddle point and then reach the attractor corresponding to cosmic acceleration at late time.

The existence of the $\phi$MDE potentially resolves the $H_0$ tension as follows.
The $H_0$ tension is the discrepancy of the estimated $H_0$ from CMB \cite{Planck2015} 
and that from the local measurements of the expansion rate of the universe.
The $H_0$ from CMB data analysis which is based on $\Lambda$CDM is lower than that from local measurements by more than 3$\sigma$ \cite{Riess:2018}.
Hence, to solve the $H_0$ tension, the dynamics of the universe should be different from that for $\Lambda$CDM.
The resolutions from modification of the late-time expansion of the universe \cite{DiValentino:2016hlg, Kumar:2016zpg, DiValentino:2017iww}
are tightly constrained by baryon acoustic oscillations (BAO) \cite{Beutler:2011hx, Ross:2014qpa, Alam:2016hwk}.
Potential resolution of the $H_0$ tension is based on the modification of the dynamics of the universe during the last scattering epoch and matter domination by early dark energy \cite{Karwal:2016vyq,Mortsell:2018mfj,Poulin:18}.
In these models, the sound horizon at the last scattering is reduced and therefore the CMB acoustic peaks shift to smaller angular scales.
Then the location of the acoustic peaks can shift to the larger angular scales and match with the data when $H_0$ increases \cite{Poulin:18}.
Possible other resolutions can be found in \cite{Banerjee2020,Krishnan2021}.

For coupled dark energy models with $\phi$MDE,
a small fraction of energy density for dark energy during the $\phi$MDE rises the effective equation of state parameter $w_{\rm eff} = \Omega_\phi w_\phi = \Omega_\phi$ to slightly positive.
Here, $\Omega_\phi$ and $w_\phi$ are the density parameter and equation of state parameter of scalar-field dark energy.
The positive effective equation of state parameter during matter domination can also shift the CMB acoustic peaks to smaller angular scales leading to a higher $H_0$ \cite{Amendola:2020}.
However, the cosmic evolution from radiation domination through $\phi$MDE towards acceleration epoch cannot be achieved in the models of coupled dark energy inspired from the conformal transformation \cite{Amendola:2006}.
This sequence of evolutions can be realized in the coupled dark energy model in which the coupling term consists of $\z \equiv u^\mu\partial_\mu\phi$ \cite{Amendola:2020}.

From the above discussion,
we see that the coupled dark energy model containing  $\z$ in the coupling term could solve both $H_0$ and $\sigma_8$ tensions.
This inspires us to study the coupled dark energy model from the general conformal transformation,
in which the coefficient of the transformation depends on both the scalar field and its kinetic term.
We are interested in whether the $H_0$ and $\sigma_8$ tensions can be alleviated in this model.
Different from the cases of usual conformal and disformal couplings,
the time derivative of the density perturbations in CDM appears in the effective coupling term for this model of coupled dark energy. 
Hence, this could   differently affect the growth of matter perturbations on small scales.

This paper is organized as follows.
In Sec.~\ref{sec2} we study the coupling between dark energy and dark matter inspired by the general conformal transformation in which the conformal coefficient depends both on scalar field and its kinetic term.
We investigate the evolution of background universe in Sec.~\ref{sec3},
and study the growth of matter perturbations on small scales in \ref{sec4}.
We give the conclusion in \ref{sec5}.

\section{Coupled dark energy model from general conformal transformation}
\label{sec2}

Let us consider the general conformal transformation defined by
\begin{equation}
\label{gbar-gen}
\bar{g}_{\mu\nu} = C(X, \phi)g_{\mu\nu} \,,
\end{equation}
where the coefficient of the conformal transformation $C$ depends on the scalar field $\phi$ and its kinetic term $X \equiv - g^{\al\bt}\pa_\al \phi \pa_\bt \phi / 2$. 
This form of the conformal transformation transforms the Einstein-Hilbert action to the action of  DHOST theories in the class where the propagation speed of gravitational waves is equal to speed of light and gravitational waves do not decay to dark energy perturbations \cite{Zumalaca,Creminelli2018}.
From the above metric transformation, we have
\ba
\bar{g}^{\mu\nu} = \frac 1{C(X, \phi)}g^{\mu\nu} \,.
\ea
In order to construct the coupled dark energy model inspired from the conformal transformation,
we suppose that the dark energy is in the form of a scalar field $\phi$ involving the conformal transformation, 
and therefore the interaction between the dark energy and the dark matter arises when the Lagrangian of the dark matter depends on the metric $\bar{g}_{\mu\nu}$ defined in Eq.~(\ref{gbar-gen}).
Hence, the model of coupled dark energy can be described by the action in which the gravitational part of the action is written in terms of the metric $g_{\mu\nu}$ while the part of the coupled matter is written in terms of $\bar{g}_{\mu\nu}$ as
\begin{equation}
S = \int d^4x \Big[\sqrt{-g}\Big(\frac{1}{2}R + P(X, \phi) + \mathcal{L}_\m(g_{\mu\nu})
\Big) + \sqrt{-\bar{g}}\mathcal{L}_\dm(\bar{g}_{\mu\nu}, \psi, \psi_{,\mu})\Big] \,,
\label{action}
\end{equation}
where we have set $1/\sqrt{8\pi G} = 1$, $R$ is the Ricci scalar, $g$ is the determinant of the metric $g_{\mu\nu}$,
$P(X, \phi) \equiv X - V(\phi)$,
$V(\phi)$ is the potential of the scalar field,
$\mathcal{L}_\m$ is the Lagrangian of ordinary matter including baryon and radiation,
 $\mathcal{L}_\dm$ is the Lagrangian of dark matter, $\psi$ is the matter field and $\psi_{, \mu}$ is the partial derivative of the field.
Varying this action with respect to $g_{\al\bt}$,
we obtain the Einstein equation in the form
\be
G^{\al\bt} = T^{\al\bt}_\phi + T^{\al\bt}_\dm + T^{\al\bt}_\m\,,
\label{eom-g1}
\ee
where $G^{\al\bt}$ is the Einstein tensor computed from $g_{\mu\nu}$,
and the energy-momentum tensors for scalar field and matter are defined in unbarred frame as
\ba
T^{\mu\nu}_{\phi} &\equiv& \frac{2}{\sqrt{-g}}\frac{\delta(\sqrt{-g}P(\phi,X))}{\delta g_{\mu\nu}}\,, 
\quad
T^{\mu\nu}_{\m} \equiv \frac{2}{\sqrt{-g}} \frac{\delta(\sqrt{-g} \mathcal{L}_\m)}{\delta g_{\mu\nu}}\,,
\label{tmn-p}\\
T^{\mu\nu}_{\dm} &\equiv& \frac{2}{\sqrt{-g}}\frac{\delta\(\sqrt{-\bar{g}}\mathcal{L}_{\dm}\)}{\delta g_{\mu\nu}}\,.
\label{tmn-m}
\ea
From these definitions of the energy-momentum tensor and $\nabla_\al G^{\al\bt}=0$ as well as the conservation of the energy-momentum tensor for the ordinary matter,
we have $\nabla_\al (T^{\al\bt}_\phi + T^{\al\bt}_\dm) =0$.
Here, $\nabla_\al$ is the covariant derivative compatible with the metric $g_{\al\bt}$. 
However, we see that the energy-momentum tensors of dark energy and dark matter do not separately conserve
because the Lagrangian of dark matter depends on field $\phi$.
Since the metric tensor does not depends on $\psi$,
variation of the action  (\ref{action}) with respect to $\psi$ yields
\be
\bar\nabla_\al \bar{T}^{\al\bt}_\dm =0\,,
\label{conserve-tb}
\ee
where $\bar\nabla_\al$ is defined from barred metric.
This implies the conservation of  $\bar{T}^{\al\bt}_\dm$ in the barred frame.
The energy-momentum tensor in the barred frame is related to that in the unbarred frame defined in Eq.~(\ref{tmn-m}) through the relation
\be
T^{\al\bt}_\dm 
= \frac{\sqrt{-\bar{g}}}{\sqrt{-g}} \frac{\de \bar{g}_{\rho\sig}}{\de g_{\al\bt}} 
\frac{2}{\sqrt{-\bar{g}}} \frac{\delta\(\sqrt{-\bar{g}}\mathcal{L}_{\dm}\)}{\delta \bar{g}_{\rho\sig}} 
= \frac{\sqrt{-\bar{g}}}{\sqrt{-g}}\frac{\de \bar{g}_{\rho\sig}}{\de g_{\al\bt}}\bar{T}^{\rho\sig}_\dm\,.
\label{bar-unbar}
\ee
Varying the action with respect to the field $\phi$, 
we obtain the evolution equation for scalar field as
\be
\nabla_\al\nabla^\al\phi - V_{,\phi} 
+ Q = 0\,,
\label{gkg-q}
\ee
where a subscript ${}_{, \phi}$ denotes derivative with respect to the field $\phi$.
The coupling term $Q$ in the above equation is a result from a variation of the dark matter action $\int d^4x \sqrt{-\bar{g}} \mathcal{L}_\dm$ in Eq.~(\ref{action}) with respect to $\phi$.
The variation of this part of the action can be computed as
\be
\delta\int d^4x \sqrt{-\bar{g}} \mathcal{L}_\dm
=
\int d^4x \de\phi \left\{
\frac{\sqrt{-\bar{g}}}2 \bar{T}^{\al\bt}_\dm\Cp g_{\al\bt}
+ \frac 12 \nabla_\sig \(\sqrt{-\bar{g}}\bar{T}^{\al\bt}_\dm g_{\al\bt} \CX \phi^{,\sig}\) 
\right\}\,,
\label{dsmg}
\ee
where the subscript ${}_{, X}$ denotes derivative with respect to $X$ and $\phi^{,\sigma}  \equiv \partial^\sigma\phi$.
Using Eq.~(\ref{bar-unbar}),
we have
\be
\sqrt{-g}T^{\al\bt}_\dm
=
C\sqrt{-\bar{g}} \bar{T}^{\al\bt}_\dm
- \frac 12 \CX \phi^{,\al} \phi^{,\bt} \sqrt{-\bar{g}} 
g_{\rho\sig}\bar{T}^{\rho\sig}_\dm\,.
\ee
Applying $g_{\al\bt}$ to the both side of the above equation, and  setting $T_\dm \equiv g_{\al\bt} T^{\al\bt}_\dm$,
we can write the above equation as
\be
\sqrt{-g}T_\dm
=
\(C + \CX X\) \sqrt{-\bar{g}} g_{\rho\sig}\bar{T}^{\rho\sig}_\dm\,,
\ee
which yields
 \be
\sqrt{-\bar{g}}g_{\al\bt} \bar{T}^{\al\bt}_\dm
=
\frac{\sqrt{-g}T_\dm}
{C + \CX X }\,.
\ee
Using this relation,
we can write Eq.~(\ref{dsmg}) as
\be
\delta\int d^4x \sqrt{-\bar{g}} \mathcal{L}_\dm 
=
\int d^4x \sqrt{-g}\,\de\phi \left\{
\frac{\Cp}{2 \(C + \CX X\)} T_\dm
+ \frac 12 \nabla_\bt \( \frac{\CX}{C + \CX X} \phi^{,\bt} T_\dm \) 
\right\}\,.
\label{dsmg1}
\ee
Combining this equation with Eq.~(\ref{gkg-q}),
we get
\be
\nabla_\al\nabla^\al\phi - V_{,\phi} 
=
-  \tCp T_\dm
- \nabla_\bt \( \tCX \phi^{,\bt} T_\dm \)
\equiv -Q\,,
\label{gkg-qg}
\ee
where $\tCp \equiv \Cp / [2 (C + \CX X)]$and $\tCX \equiv \CX /[2(C + \CX X)]$.
Eq.~(\ref{gkg-qg}) can be written in terms of the energy-momentum tensor as
\be
\nabla_{\al}T^\al_\bt{}_\phi
=
 -Q \phi_{,\bt}\,, 
\label{grdtphi}
\ee
where $T^\al_\bt{}_\phi$ is the energy-momentum tensor of the scalar field.
According to the conservation of the total energy-momentum tensor,
the above equation gives
\be
\nabla_{\al}T^\al_\bt{}_\dm
=
 Q \phi_{,\bt}\,.
\label{grdmatt}
\ee
In the case of a conformal transformation in which the conformal coefficient $C$ depends only on the field $\phi$, 
we have $\CX = 0$ and therefore Eq.~(\ref{gkg-qg}) reduces to the equation for the case of usual conformal transformation.
When $\CX$ is not vanish, the coupling term $Q$ contains coupling between the field derivative and the energy density as well as between the field derivative and the derivative of energy density of CDM. 
The latter form of the coupling can lead to different effects on cosmic evolution and the growth of matter perturbations compared with the usual conformal coupling case.

\section{Evolution of the background Universe}
\label{sec3}

In this section, we study effects of the interaction between dark energy and dark matter due to the general conformal transformation on the evolution of the background universe.
Using the Friedmann-Lema\^{i}tre-Robertson-Walker (FLRW) metric,
\be
ds^2 = - dt^2  + a^2 \delta_{ij} dx^i dx^j\,,
\ee
where $\delta_{ij}$ is the Kronecker delta, and supposing that the scalar field is homogeneous and other matter components in the universe are described by perfect fluid,
Eqs.~(\ref{gkg-qg}) and (\ref{grdmatt}) become
\be
\ddot\phi + 3H\dot\phi + V_{,\phi} 
= \bar{Q}\,,
\label{gkg-qgb}
\ee
and
\be
\dot\rho_\dm + 3H\rho_\dm
= - \bar{Q} \dot\phi\,.
\label{rhomdot}
\ee
Here, $\rho_\dm$ is the energy density of dark matter, $H \equiv \dot a / a$ is the Hubble parameter, a dot denotes a derivative with respect to time $t$, and 
\begin{equation}
\bar{Q} = - \tCp \rho_\dm + \(\ddot\phi + 3H\dot\phi\)\tCX \rho_\dm + 2 \tCXp X \rho_\dm 
+ 2 \tCXX \ddot\phi X \rho_\dm + \tCX \dot\phi \dot\rho_\dm \,.
\label{qbar}
\end{equation}
We see that the interaction term $\bar{Q}$ in the above equation depends on $\ddot\phi$ and $\dot\rho_\dm$.
Hence, we combine Eqs.~(\ref{gkg-qgb}) and (\ref{rhomdot}) to write the evolution equations for $\phi$ and $\rho_\dm$ in the forms
\be
\ddot\phi + 3 H \dot\phi +V_{, \phi} = Q_0\,,
\quad\mbox{and}\quad
\dot\rho_\dm + 3 H \rho_\dm = - \dot\phi Q_0\,.
\label{ddpdrq0}
\ee 
Here, we define the effective coupling term as
\be
\tQ \equiv \frac{Q_0}{\rho_\dm}
=
\frac{\Theta V_{, \phi } + 3 H \Theta \dot{\phi } -2 X \tCXp + \tCp}{\Theta\rho_\dm -2 X \tCX - 1}\,,
\label{tq}
\ee
where  $\Theta \equiv \tCX +2 X \tCXX$.
Since the energy-momentum tensors of baryon and radiation are separately conserve,
in the background universe the conservation of these energy-momentum tensors yields
\be
\dot\rho_b = - 3 H \rho_b\,,
\quad\mbox{and}\quad
\dot\rho_r = - 4 H \rho_r\,,
\label{drbdrr}
\ee
where $\rho_b$ and $\rho_r$ are the energy density of baryon and radiation.

\subsection{Autonomous equations}
\label{sec:3.1}

Let us compute the autonomous equations by defining the dimensionless dynamical variables as
\ba
x &=& \frac{\dot\phi}{\sqrt{6}H}, \quad y = \frac{V}{3 H^2}, \quad \Omega_\dm = \frac{\rho_\dm}{3 H^2}\,,
\nonumber\\
\Omega_b &=& \frac{\rho_b}{3 H^2}\,, \quad \Omega_r = \frac{\rho_r}{3 H^2}\,,
\ea
and the dimensionless functions as
\begin{align}
z &= \frac{\CX}{C} H^2\,, \quad \lambda = \frac{V_{, \phi}}{V}\,, \\ 
\gamma &= \tCp\,, \quad \chi = \tCX H^2 \,.
\label{def:fn}
\end{align}
In terms of the above dimensionless variables, 
the Friedmann equation gives
\be
1 = x^2 + y + \Omega_\dm + \Omega_b + \Omega_r\,.
\label{friedmann}
\ee
From the above dimensionless variables, we obtain autonomous equations from Eq.~(\ref{ddpdrq0}) as 
\begin{align}
\label{xp}
x' =& -x\frac{\dot{H}}{H^2} \nonumber\\  
&+ \frac{ 
\sqrt{6} \left(\gamma \Omega_\dm + \lambda y\right) + 6 x + 3 \sqrt{6} \left(\gamma z \Omega_\dm - 2 \Omega_\dm \chi_{,\phi} + 2 \lambda y z\right) x^2
+ 36 z x^3 - 18 \sqrt{6} z \Omega_\dm \chi_{,\phi} x^4}{36 x^2 \Omega_\dm \left(3 x^2 z + 1\right) \chi_{,X} + 3 z \Omega_\dm - 12 x^2 z - 2}\,,
\\
\label{yp}
y' &= \sqrt{6} \lambda x y - 2 y \frac{\dot{H}}{H^2}\,,
\\
\label{zp}
z' &= 6 x \left(\frac{\CX}{C}\right)_{, X} \left(x \frac{\dot{H}}{H^2} + x'\right) + 
\sqrt{6} x \left(\frac{\CX}{C}\right)_{, \phi} + 2 z\frac{\dot{H}}{H^2}\,, 
\\
\Omega_\dm' =& - 2 \Omega_\dm\frac{\dot{H}}{H^2}
-\frac{36 x^3 (3 x^2 z + 1) (6 x + \sqrt{6} \lambda y) \chi_{,X} + 3 \sqrt{6} \lambda x y z - 2 \sqrt{6} \gamma x (3 x^2 z + 1)}{36 x^2 \Omega_\dm \left(3 x^2 z + 1\right) \chi{,_X} + 3 z \Omega_\dm - 12 x^2 z - 2}\Omega_\dm
\nonumber\\
\label{omp}
&
+\frac{12 \sqrt{6} x^3 \left(3 x^2 z + 1\right) \chi_{,\phi} + 18 x^2 z + 6}{36 x^2 \Omega_\dm \left(3 x^2 z + 1\right) \chi_{,X} + 3 z \Omega_\dm - 12 x^2 z - 2}\Omega_\dm
\nonumber\\ 
&-\frac{9 \big[12 \left(3 x^4 z + x^2\right) \chi_{,X} + z\big]}{36 x^2 \Omega_\dm \left(3 x^2 z + 1\right) \chi_{,X} + 3 z \Omega_\dm - 12 x^2 z - 2}\Omega_\dm^2\,,
\end{align} 
where a prime denotes a derivative with respect to $N \equiv \ln a$ and
\begin{align}
\label{dhfull}
\frac{\dot{H}}{H^2} = \frac{1}{2} \left(\Omega_\dm - 2 x^2 + 4 y - 4\right)\,.
\end{align}
From Eq.~(\ref{drbdrr}), 
we get
\ba
\Omega_b' &=& - 3 \Omega_b - 2 \frac{\dot{H}}{H^2}\Omega_b\,,
\label{ombp}\\
\Omega_r' &=& - 4 \Omega_r - 2 \frac{\dot{H}}{H^2}\Omega_r\,.
\label{omrp}
\ea
Let us consider the denominator of the terms in Eqs.~(\ref{xp}) and (\ref{omp}).
The denominators of all terms except the terms which are proportional to $\dot H / H^2$ are the same and can vanish when
\be
\chi_{,X} = - \frac{3 z \Omega _\dm-12 x^2 z-2}{36 x^2 \Omega _\dm \left(3 x^2 z+1\right) }\,.
\label{d0}
\ee
This suggests that $x'$ and $\Omega_\dm'$ can be infinite when the above equation is satisfied.
To ensure that the background universe properly evolves,
the situations in which the above equation is satisfied have to be avoided.

To perform further analysis,
we use the coefficient $C$ in the simple form that could reveal main features of this form of coupling.
Since the coupling terms in the evolution equations depend on derivatives of $C$ with respect to $\phi$ and $X$,
the coefficient $C$ should be a polynomial function of $X$.
Moreover, the following autonomous equations can be a complete set of equations if the coupling term $\tQ$ given in Eq.~(\ref{tq}) does not depend on $\phi$.
Hence, we choose the potential of the scalar field and coefficient $C$ in the forms 
\begin{equation}
V(\phi) = V_0 \e^{\lambda \phi}\,,
\quad 
C(\phi, X) = C_0 \e^{\lambda_1\phi}\left[1 + \e^{\lambda_2\phi} \left(\frac{X}{\Lambda_0}\right)^{\lambda_3}\right] \,,
\label{CoefC}
\end{equation}
where $C_0, \lambda_1, \lambda_2$ and $\lambda_3$ are dimensionless constants,
while $V_0$ and $\Lambda_0$ are constants with the same dimension as $X$.
According to Eq.~(\ref{def:fn}),
the above form of the potential implies that $\lambda$ is a dimensionless constant.
For this form of $C$,
Eq.~(\ref{d0}) gives
\be
\lambda_3 = \frac{27 x^2 z^2 \Omega _\dm+3 z \Omega _\dm+36 x^4 z^2+18 x^2 z+2}{6 z \Omega _\dm}\,.
\label{l3sing}
\ee
For the case of positive $\lambda_3$,
we get $z > 0$ according to the definition in Eq.~(\ref{def:fn}).
This suggests that the above equation can be satisfied if $\lambda_3 > 0$.
This implies that $x'$ and $\Omega_\dm'$ can be infinite at some time during the evolution of the universe
if $\lambda_3$ is positive.
Based on the numerical investigation,
the divergence of $x'$ and $\Omega_\dm'$ can be avoided if $\lm_3 < 1$.

\subsection{Fixed points}

Since we are interested in the fixed points corresponding to the matter dominated epoch and the late-time universe,
we ignore the contribution from the radiation energy density in the dynamical analysis.
To compute the fixed points, we also drop the contribution from baryon  because Eq.~(\ref{ombp}) has fixed points at $\Omega_b = 0$ and at $\dot{H} / H^2 = -3/2$.
The first point can be reached in the future while the second point corresponds to the matter dominated epoch.
The second point is not exactly compatible with $\phi$MDE because the $\phi$MDE requires $\dot{H} / H^2 = -3 (1 + w_{\rm eff})/2 \lesssim -3/2$ during matter domination.
Hence, to study the $\phi$MDE point in the dynamical analysis,
we drop the contribution from the baryon energy density.
However, it will be shown in the numerical integration that the inclusion of baryon energy density does not forbid the existence of $\phi$MDE,
because we still get $\Omega_b' \sim 0$ when $\dot{H} / H^2 \lesssim -3/2$.

Ignoring the contributions from radiation and baryon energy density,
Eq.~(\ref{friedmann}) gives
\be
\Omega_\dm = 1 - x^2 - y\,.
\label{omcfried}
\ee
Substituting this expression into Eq.~(\ref{dhfull}),
we get
\be
\label{hdot1}
\frac{\dot{H}}{H^2} = -\frac{3}{2} \left(x^2 - y + 1\right)\,.
\ee 
Setting $y'=0$, Eq. (\ref{yp}) is satisfied by two solutions which correspond to the fixed points $y_c = 0$ and 
\be
\label{hdot2}
\frac{\dot{H}}{H^2} = \sqrt{\frac{3}{2}} \lambda x_c \,,
\ee
where the subscript ${}_c$ denote evolution at the fixed point.

\subsubsection{Field dominated point and scaling point}

We first consider the fixed point $y_c \neq 0$.
We can compute $y$ at the fixed point by matching Eq.~(\ref{hdot1}) with Eq.~(\ref{hdot2})  which yields 
\be
\label{yc}
y_c = \sqrt{\frac{2}{3}} \lambda  x_c + x_c^2 + 1\,.
\ee
Substituting Eqs.~(\ref{omcfried}), (\ref{hdot2}) and (\ref{yc}) into Eqs.~(\ref{xp}) and (\ref{zp}),
then  inserting $C$ from Eq.~(\ref{CoefC}) into the resulting equations, 
and finally setting $x'=z'=0$, we obtain the equations for the fixed points,
\begin{align}
\label{e2b}
0 &= 
-\sqrt{6} \lambda \lambda_3 
+ \left(\lambda \lambda_1 - 2 \left(\lambda^2 + 3\right)\right) \lambda_3 x_c
+ \sqrt{6} \lambda_3 \left(\lambda_1 - \lambda (9 z_c + 2)\right) x_c^2 
\nonumber\\
& + 3 \left(-2 \lambda ^2 \lambda_3^2 + \left(-5 \lambda^2 + \lambda_1 \lambda - 2 \lambda_2 \lambda - 18\right) \lambda_3 + \lambda \lambda_2\right) z_c x_c^3 
\nonumber\\
&+ 3 \sqrt{6} z_c \left(-2 \lambda_3 \lambda_2+\lambda_2 + \lambda_3 \left(\lambda_1 -\lambda \left(2 \lambda_3 + 6 z_c + 5\right)\right)\right) x_c^4
- 9 \left(\left(\lambda^2 + 12\right) \lambda_3 - 3 \lambda \lambda_2\right) z_c^2 x_c^5
\nonumber\\
&+ 9 \sqrt{6} \left(3 \lambda_2 - \lambda \lambda_3\right) z_c^2 x_c^6
\,,\\
\label{e3b}
0 &= \frac{\sqrt{6}}{\lambda _3} \left(\lambda_2 + \lambda \lambda_3\right) x_c z_c \left(\lambda_3 - 3 x_c^2 z_c\right)\,.
\end{align}
From Eq.~(\ref{e3b}), we can solve for $z_c$ as 
\be
z_c = 0 \quad \mbox{and}  \quad z_c = \frac{\lambda _3}{3 x_c^2} \,.
\label{zc}
\ee
We concentrate on the second solution rather than  $z_c = 0$ solution,
because the $z_c = 0$ solution corresponds to the case where the kinetic dependence of $C$ is negligible, i.e. $z = \CX/C = 0$.
In addition to the above solutions,
Eq.~(\ref{e3b}) is also satisfied by the condition $\lambda _2+\lambda  \lambda _3 = 0$.
However, this case can be viewed as a special case of solutions in Eq.~(\ref{zc}),
so that we will not discuss this case in detail.

Inserting the second fixed point  of $z$ from the above equation into Eq.~(\ref{e2b}),
we obtain two fixed points of variable $x$ as
\be
x_{c} = \left\{-\frac{\lambda }{\sqrt{6}},\frac{\sqrt{6} \left(2 \lambda _3+1\right)}{\lambda _1+\lambda _2-\lambda  \left(3 \lambda _3+2\right)}\right\}\,.
\label{xcscaling}
\ee
Inserting $x_c$ from above equation into Eq. (\ref{yc}),
we obtain
\be
y_{c} = \left\{1-\frac{\lambda ^2}{6},
 1 +\frac{6 \left(2 \lambda _3+1\right)^2}{\left(\lambda _1+\lambda _2-\lambda  \left(3 \lambda _3+2\right)\right)^2}
+\frac{2 \lambda  \left(2 \lambda _3+1\right)}{\lambda _1+\lambda _2-\lambda \left(3 \lambda _3+2\right)}\right\}\,.
\label{ycscaling}
\ee
From $x_c$ and $y_c$, we can compute the density parameter and equation of state of scalar field at the fixed points from the definitions $\Omega_\phi  \equiv x^2 + y$ and $w_\phi \equiv (x^2 - y) / \Omega_\phi$ as
\ba
\label{omp:sc}
\Omega_{\phi c} &=& \left\{1, \frac{12 \left(2 \lambda _3+1\right)^2}{\left(\lambda _1+\lambda _2-\lambda  \left(3 \lambda _3+2\right)\right)^2}+\frac{2 \lambda  \left(2 \lambda _3+1\right)}{\lambda _1+\lambda _2-\lambda  \left(3 \lambda _3+2\right)}+1\right\}\,,
\\
\label{w:sc}
w_{\phi c} &=& \left\{\frac{1}{3} \left(\lambda ^2-3\right),-\frac{\lambda _1+\lambda _2+\lambda  \lambda _3}{\left(\lambda _1+\lambda _2-\lambda  \left(3 \lambda _3+2\right)\right) \left(\frac{12 \left(2 \lambda _3+1\right)^2}{\left(\lambda _1+\lambda _2-\lambda  \left(3 \lambda _3+2\right)\right)^2}+\frac{2 \lambda  \left(2 \lambda _3+1\right)}{\lambda _1+\lambda _2-\lambda  \left(3 \lambda _3+2\right)}+1\right)}\right\}\,. 
\ea
We see that the first pair of $(x_c, y_c)$ corresponds to the field dominated point, while the second pair corresponds to the scaling point.
From the above equations, we can write $\lambda$ in terms of $w_{\phi c}$ for the case of field dominated point as
\be
\lambda = \sqrt{3 (w_{\phi c} + 1)} \,,
\label{lm:fd}
\ee
which is the same as that for the field-dominated solution for uncoupled quintessence with exponential potential.
For the scaling point, we write $\lambda$ and $\lambda_1$ in terms of $\Omega_{\phi c}$ and $w_{\phi c}$ by solving Eq~(\ref{omp:sc}) and (\ref{w:sc}) for $\lambda$ and $\lambda_1$.
The results are
\begin{align}
\lambda &= \mp \frac{\sqrt{3}(w_{\phi c} \Omega_{\phi c} + 1)}{\sqrt{\left(w_{\phi  c}+1\right) \Omega _{\phi  c}}} \,,
\label{lmscaling}\\
\lambda_1 &=
-\lambda _2 \pm \frac{\sqrt{3} \left(-3 \lambda _3 w_{\phi  c} \Omega _{\phi  c}-2 w_{\phi  c} \Omega _{\phi  c}+\lambda _3\right)}{\sqrt{\left(w_{\phi  c}+1\right) \Omega _{\phi  c}}}\,.
\label{l1scaling}
\end{align}
Using the above equations,
we can compute the values of $\lambda$ and $\lambda_1$ from $\lambda_2$, $\lambda_3$, $w_{\phi c}$ and $\Omega_{\phi c}$.
 The values of $w_{\phi c}$ and $\Omega_{\phi c}$ can be specified based on observational constraints, i.e.,  
if we suppose that the scaling point corresponds to the late-time universe,
 we can set $w_{\phi c} = -0.99$ and $\Omega_{\phi c} = 0.7$.
This suggests that to perform further analysis,
we need to specify only the parameters $\lambda_2$ and $\lambda_3$ instead of all parameters of the model $\lambda$, $\lambda_1$, $\lambda_2$ and $\lambda_3$.
As a result,
the cases where the fixed points do not satisfy the observational constraints can be excluded in our analysis.
Inserting $\lambda$ and $\lambda_1$ from the above equations into Eqs. (\ref{xcscaling}) and (\ref{ycscaling}),
we obtain
\ba
x_{c} = \pm \sqrt{\frac{1}{2}\Omega_{\phi c}\left( 1 + w_{\phi c}\right)} \quad \mbox{and}
\quad
y_{c} = \frac{1}{2} \Omega_{\phi c} \left( 1 - w_{\phi c} \right) \,.
\label{xyscaling}
\ea

\subsubsection{Kinetic dominated point and $\phi$MDE point}

We now consider the point $y_c = 0$.
For this point, Eq.~(\ref{hdot1}) gives
\be
\label{hdot1y0}
\frac{\dot{H}}{H^2} = -\frac{3}{2} \left(x_c^2 + 1\right)\,.
\ee
Inserting $y_c = 0$ and Eq.~(\ref{hdot1y0}) into Eqs.~(\ref{xp}) and (\ref{zp}) 
and performing the same procedures as those for Eqs.~(\ref{e2b}) and (\ref{e3b}), we obtain
\ba
\label{e2y0}
0 &=& \(1 - x_c^2\)\[
\sqrt{6} \lambda _1 \lambda _3 
+ 3 \lambda _3 \left(\left(6 \lambda _3-3\right) z_c +2\right) x_c
+ 3 \sqrt{6} \left(-2 \lambda _3 \lambda _2+\lambda _2+\lambda _1 \lambda _3\right) z_c x_c^2 
\right.
\nonumber\\
& & \left.
 + 9 \lambda _3 z_c \left(2 \lambda _3-9 z_c + 5\right) x_c^3
+ 27 \sqrt{6} \lambda _2 z_c^2 x_c^4
+ 27 \lambda _3 z_c^2 x_c^5
\] \,, 
\\
\label{e3y0}
0 &=& -\frac{1}{\lambda _3} z_c \left(3 \lambda _3 \left(x_c^2+1\right)-\sqrt{6} \lambda _2 x_c\right) \left(\lambda _3-3 x_c^2 z_c\right)\,.
\ea
In the following consideration,
we use a superscript ${}^\fy$ to denote the quantities corresponding to the fixed point $y_c = 0$,
which will be seen in the subsequent considerations that this point  can play a role of $\phi$MDE.
From Eq.~(\ref{e3y0}), we obtain $z$ at the fixed point as
\be
z_c^\fy = 0 \quad \mbox{and} \quad
z_c^\fy = \frac{\lambda _3}{3 (x_c^\fy)^2} \,,
\ee
which are similar to the case of scaling point.
Substituting the second solution for $z_c^\fy$ into Eq.~(\ref{e3y0}),
we can solve for $x_c^\fy$ as
\be
x_{c}^{\rm kinetic} = \pm 1\,,
\quad\mbox{and}
\quad
x_c^\fy = 
-\frac{\lambda _1+\lambda _2}{\sqrt{6} \left(3 \lambda _3+2\right)} \mp \frac{\sqrt{\lambda _1^2+2 \lambda _2 \lambda _1+\lambda _2^2+6 \lambda _3 \left(3 \lambda _3+2\right)}}{\sqrt{6} \left(3 \lambda _3+2\right)}\,.
\label{xc:y0}
\ee
The first two solutions are kinetic-dominated points, while the other solutions correspond to $\phi$MDE points.
We insert $x_c^\fy$ into the definition of $\Omega_{\phi}$,
we get the expression for $\Omega_\phi$ at $y_c = 0$ in the form
\begin{align}
\Omega_{\phi c}^\fy =& \Bigg[1, 1, \frac{\left(\lambda_1 + \lambda_2 + \sqrt{\lambda_1^2 + 2 \lambda_2 \lambda_1 + \lambda_2^2 + 6 \lambda_3 \left(3 \lambda_3 + 2\right)}\right)^2}{6(3 \lambda_3 + 2)^2},\nonumber\\ 
&\frac{\left(\lambda_1 + \lambda_2 - \sqrt{\lambda_1^2 + 2 \lambda_2 \lambda_1 +\lambda_2^2 + 6 \lambda_3 \left(3 \lambda_3 + 2\right)}\right)^2}{6 (3 \lambda_3 + 2)^2}\Bigg]\,. 
\label{omp:y0}
\end{align}
Since $y = 0$ at these fixed points,
we get $w_{\phi c}^\fy = 1 $.
Hence, the effective equation of state parameter $w_{\rm eff}  = \Omega_\phi w_\phi = \Omega_{\phi c}^\fy$ is slightly positive during the $\phi$MDE.
Similar to scaling fixed point, 
  we write $\lambda_1$ in terms of $\Omega_{\phi c}^\fy ,\lambda_2$ and $\lambda_3$ using Eq.~(\ref{omp:y0}) as
\ba
\lambda_{1}^\fy = -\lambda _2 \mp \sqrt{\frac{3}{2}} \frac{\left|3 \lambda _3 \Omega _{\phi  c}^\fy + 2 \Omega _{\phi  c}^\fy - \lambda _3\right|}{\sqrt{\Omega _{\phi  c}^\fy}}\,.
\label{l1:y0}
\ea
In the following consideration, we use the subscripts ${}_-$ and ${}_+$ to indicate the selected sign in the expressions which contain $\pm$ or $\mp$.
As an example, if we apply this notation to Eq.~(\ref{xc:y0}),
we get
\be
x_{c+}^\fy = 
-\frac{\lambda _1+\lambda _2}{\sqrt{6} \left(3 \lambda _3+2\right)} + \frac{\sqrt{\lambda _1^2+2 \lambda _2 \lambda _1+\lambda _2^2+6 \lambda _3 \left(3 \lambda _3+2\right)}}{\sqrt{6} \left(3 \lambda _3+2\right)}\,.
\ee
Using such notation, the possible expressions of $\lambda$ and $\lambda_1$ for the scaling points can be expressed as follows:
according to Eqs.~(\ref{lmscaling}) and (\ref{l1scaling}),
there are two possible forms of $\lambda$ and $\lambda_1$ such that $(\lambda, \lambda_1) =  (\lambda_-, \lambda_{1+})$ and $(\lambda_+, \lambda_{1-})$.
For $\phi$MDE point, Eq.~(\ref{xc:y0}) shows that there are two possible forms of $x_{c}^\fy$, i.e., $x_{c-}^\fy$ and $x_{c+}^\fy$.
Each of them leads to two possible choices of $\lambda_1$ given in Eq. (\ref{l1:y0}).

\subsection{Stability}

We now consider stability of the fixed points considered in the previous section by linearizing the autonomous equations (\ref{xp}) - (\ref{zp}) around the fixed points.
Before performing the linearization,
we set $\Omega_\dm = 1 - x^2 - y$ and use $C$ from Eq.~(\ref{CoefC}).
The linearized equations can be written in the matrix form, 
and the stability of the deviation around the fixed points can be estimated from the signs of the eigenvalues of the Jacobian matrix defined by
\be
J_{ij} = \left. \frac{\partial x_i'}{\partial x_j}\right|_{\rm fixed \, point}\,,
\ee
where $x_i = (x,y, z)$.

\subsubsection{Field dominated point}

We first consider the field dominated point in which $x$ and $y$ at fixed point are given by the first solution in Eqs.~(\ref{xcscaling}) and (\ref{ycscaling}),
 while $z$ at the fixed point is the second solution in Eq.~(\ref{zc}).
The eigenvalues for this case are
\ba
\mu_1 &=&
3 \lambda_3 (1 + w_{\phi c}) + \lambda _2 \sqrt{3 (1 + w_{\phi c})} \,, 
\nonumber\\
\mu_2 &=& -\frac{3}{2} \left(1 - w_{\phi c}\right)\,,
\nonumber\\
\mu_3 &=& \frac{\lambda _3 \left(9 w_{\phi  c}-3\right)+6 w_{\phi  c} - \sqrt{3} \left(\lambda _1+\lambda _2\right) \sqrt{1 + w_{\phi  c}}}{4 \lambda _3+2}\,,
\ea
where we have expressed $\lambda$ in terms of $w_\phi$ at the fixed point using Eq.~(\ref{lm:fd}).
One can check that the field dominated point is stable when both of the following conditions are satisfied
\ba
\lambda_3 & < & -\frac{\lambda_2}{\sqrt{3\sigma_1}}\,,
\label{l3fd}\\
\lambda_1 & & \left \{\begin{array}{cc}
<  -\frac{2 w_{\phi  c} \left(2 \sqrt{3} \lambda _2-3 \sqrt{w_{\phi  c}+1}\right)}{\sqrt{3} \left(w_{\phi  c}+1\right)}
& \mbox{for} \,\, \lambda_3 < -1/2 \\
>  -\frac{2 w_{\phi  c} \left(2 \sqrt{3} \lambda _2-3 \sqrt{w_{\phi  c}+1}\right)}{\sqrt{3} \left(w_{\phi  c}+1\right)}
& \mbox{for} \,\, \lambda_3 > -1/2 \\
\end{array} \right. \,.
\label{l1fd}
\ea
Since $\mu_2$ is always negative when $w_{\phi c} < 1$ which is the case for scalar field with standard kinetic term,
the field dominated points cannot  be unstable.

\subsubsection{Scaling fixed point}

\begin{figure}
\includegraphics[height=0.4\textwidth, width=0.9\textwidth,angle=0]{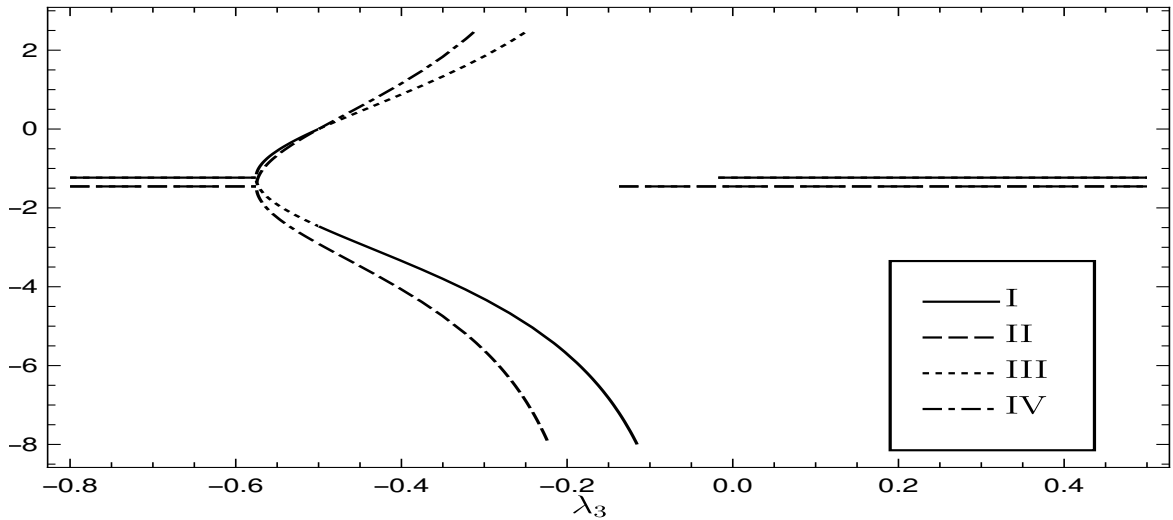}
\caption{\label{fig1.eps}
Plots of the real parts of $\mu_2$ and $\mu_3$ for scaling fixed point.
In the plots, $w_{\phi c} = -0.99$ and $\lambda_2 = 1$.
The lines I and II represent the real part of $\mu_2$ while the
lines III and IV represent the real part of $\mu_3$.
The lines I and III show the cases of $\Omega_{\phi c} = 0.65$ while the
lines II and IV show the cases of $\Omega_{\phi c} = 0.95$.
}
\end{figure}

For the scaling point in which the expressions for $x_c$ and $y_c$ are given in Eq. (\ref{xyscaling}),
the eigenvalues are
\ba
\mu_1 &=&  3 \lambda_3 \left(1 + w_{\phi c} \Omega_{\phi c} \right) \mp \lambda_2 \sqrt{3 \Omega_{\phi c} (1 + w_{\phi c})} \,,
\nonumber\\
\mu_2 &=& -\frac{3}{4} \left(1 - w_{\phi  c} \Omega_{\phi  c} \right) + 3 \sqrt{\frac{r_a}{r_b}}\,,
\quad\mbox{and}\quad
\mu_3 = -\frac{3}{4} \left(1 - w_{\phi c} \Omega_{\phi c} \right) - 3 \sqrt{\frac{r_a}{r_b}} \,,
\label{mu23sl}
\ea
where
\ba
r_a &=&  \lambda_3 \left(w_{\phi c}^2 \left(2 w_{\phi c} + 1\right) \Omega_{\phi c}^3 + \left(-3 w_{\phi c}^2 - 18 w_{\phi c} + 16\right) \Omega_{\phi c}^2 + \left(16 w_{\phi c} - 15\right) \Omega_{\phi c} + 1\right) 
\nonumber\\
&& + \Omega_{\phi c} \left(w_{\phi c}^2 \left(w_{\phi c} + 1\right) \Omega_{\phi c}^2 - 2 \left(w_{\phi c}^2 + 5 w_{\phi c} - 4\right) \Omega_{\phi c} + 9 w_{\phi c} - 7\right)\,,
\label{ra:def}\\
r_b &=& 16 \left(\lambda_3 \Omega_{\phi c} + 2 \lambda_3 w_{\phi c} \Omega_{\phi c} + w_{\phi c} \Omega_{\phi c} + \Omega_{\phi c} + \lambda_3\right)\,.
\label{rb:def}
\ea
In the above eigenvalues,
we have written $\lambda$ and $\lambda_1$ in terms of $w_{\phi c}$ and $\Omega_{\phi c}$ using Eqs.~(\ref{lmscaling}) and (\ref{l1scaling}).
The fixed point $x_{c +}$ and $x_{c -}$ in Eq. (\ref{xyscaling}) lead to the same $\mu_2$ and $\mu_3$ but different $\mu_1$.
The first eigenvalue can be negative when 
\be
\lambda_3 < \pm \frac{ \lambda_2 \sqrt{3 \Omega_{\phi c} (1 + w_{\phi c}) }}{3 (1 + w_{\phi c}\Omega_{\phi c} )} \,.
\label{l31sc}
\ee
The eigenvalues $\mu_2$ and $\mu_3$ in Eq.~(\ref{mu23sl}) can be infinite if $r_b = 0$ which occurs when
\be
\lambda_3 = \lambda_{3 b} = -\frac{\left(w_{\phi  c}+1\right) \Omega _{\phi  c}}{2 w_{\phi  c} \Omega _{\phi  c}+\Omega _{\phi  c}+1}\,.
\ee
The real parts of both $\mu_2$ and $\mu_3$ can be ensured to be negative if the ratio $r_a/r_b < 0$.
To check the sign of this ratio, we also compute $\lambda_3$ at which $r_a =0$.
It can be shown that $r_a = 0$ when
\be
\lambda_3 = \lambda_{3 a} =
-\frac{\Omega _{\phi  c} \left(w_{\phi  c}^2 \left(w_{\phi  c}+1\right) \Omega _{\phi  c}^2-2 \left(w_{\phi  c}^2+5 w_{\phi  c}-4\right) \Omega _{\phi  c}+9 w_{\phi  c}-7\right)}
{w_{\phi  c}^2 \left(2 w_{\phi  c}+1\right) \Omega _{\phi  c}^3+\left(-3 w_{\phi  c}^2-18 w_{\phi  c}+16\right) \Omega _{\phi  c}^2+\left(16 w_{\phi  c}-15\right) \Omega _{\phi  c}+1}\,.
\ee
For $\Omega_{\phi c} > 0.6$ and $w_{\phi c} \gtrsim -1$,
the coefficient of $\lambda_3$ in Eq.~(\ref{ra:def}) is negative while that in Eq.~(\ref{rb:def}) is positive.
Hence, $r_b < 0$ when $\lambda_3 < \lambda_{3 b}$ while $r_a < 0$ when $\lambda_3 > \lambda_{3 a}$.
Since $\lambda_{3 a} < \lambda_{3 b}$, the ratio $r_a/r_b$ is negative when  $\lambda_3 < \lambda_{3 a}$ or $\lambda_3 > \lambda_{3 b}$.
As a result, the scaling point is stable when  $\lambda_3 < \lambda_{3 a}$ or $\lambda_3 > \lambda_{3 b}$ for suitable choice of $\lambda_2$ according to Eq.~(\ref{l31sc}).
For the case $\lambda_3 \in (\lambda_{3 a}, \lambda_{3 b})$,
we have to evaluate $\mu_2$ and $\mu_3$ numerically.
The real parts of $\mu_2$ and $\mu_3$ for some choices of $\Omega_{\phi c}$ are plotted in Fig.~\ref{fig1.eps}.
From this figure,
the real parts of the eigenvalues weakly depend on $\lambda_2$.

\subsubsection{Kinetic dominated point and $\phi$MDE point}

We first consider the kinetic dominated points where $x_c = \pm 1$.
For these points,
the eigenvalues are
\be
\mu_1 = \frac{3 \left(\lambda _3+1\right)}{2 \lambda _3+1} \pm \frac{\sqrt{6} \left(\lambda _1+\lambda _2\right)}{4 \lambda _3+2}\,,
\quad
\mu_2 = 6 \lambda _3 \mp \sqrt{6} \lambda _2\,, 
\quad\mbox{and}\quad
\mu_3 = 6 \pm \sqrt{6} \lambda\,.
\label{mu:kin}
\ee
The second eigenvalue $\mu_2$ can be either positive or negative depending on the values of $\lambda_2$ and $\lambda_3$.
This means that these kinetic points can be saddle point,
and therefore these points could be reached for some ranges of $\lambda_2$, $\lambda_3$ and some choices of initial conditions.
However, we are interested in the cases where the cosmic evolution satisfies observational data,
so that we will not discuss these points in more detail.

We next consider the $\phi$MDE points given by Eq.~(\ref{xc:y0}).
The eigenvalues for these fixed points are complicated and their values consist of many possible cases according to the range of $\lambda_1, \lambda_2$ and $\lambda_3$.
However, if we are interested in the case where the $\phi$MDE is followed by accelerating epoch described by scaling points,
we have to demand that $\lambda_1$ from Eq.~(\ref{l1scaling}) is equal to that from Eq.~(\ref{l1:y0}) .
Matching these two equations,
we get the relation between $\Omega_{\phi c}^\fy$ and $\Omega_{\phi c}$ as
\be
\Omega_{\phi c \mp}^{\fy} = \frac{A \mp \left| \lambda _3 \left(3 w_{\phi  c} \Omega _{\phi  c}-1\right)+2 w_{\phi  c} \Omega _{\phi  c} \right| \sqrt{B}}{\left(3 \lambda _3+2\right){}^2 \left(w_{\phi  c}+1\right) \Omega _{\phi  c}} \,,
\label{Om:y0}
\ee
where 
\ba
A &=& \lambda _3^2 \left(9 w_{\phi  c}^2 \Omega _{\phi  c}^2-3 \left(w_{\phi  c}-1\right) \Omega _{\phi  c}+1\right)+2 \lambda _3 \Omega _{\phi  c} \left(6 w_{\phi  c}^2 \Omega _{\phi  c}-w_{\phi  c}+1\right)+4 w_{\phi  c}^2 \Omega _{\phi  c}^2 \,,
\\
B &=& \lambda _3^2 \left(9 w_{\phi  c}^2 \Omega _{\phi  c}^2+6 \Omega _{\phi  c}+1\right)+4 \lambda _3 \Omega _{\phi  c} \left(3 w_{\phi  c}^2 \Omega _{\phi  c}+1\right)+4 w_{\phi  c}^2 \Omega _{\phi  c}^2\,.
\ea
The right-hand side of Eq.~(\ref{Om:y0}) could be infinite when $\lambda_3$ is equal to $-2/3$.
Nevertheless, if we take the limit $\lambda_3 \to -2/3$, Eq.~(\ref{Om:y0}) gives  
\be
\Omega_{\phi c - }^{\fy} = \frac{1}{2} \left(w_{\phi  c}+1\right) \Omega _{\phi  c} \quad, \quad \Omega_{\phi c + }^{\fy} = \infty  \,.
\ee
Hence, from now we consider only $\Omega_{\phi c - }^{\fy}$ which will be denoted by $\Omega_{\phi c }^{\fy}$.
It follows from Eq.~(\ref{Om:y0}) that  $\Omega_{\phi c }^{\fy}$ can have an imaginary part if $B$ is negative which occurs when
\be
\frac{2 \left(-3 w_{\phi  c}^2 \Omega _{\phi  c}^2-\sqrt{\Omega _{\phi  c}^2-w_{\phi  c}^2 \Omega _{\phi  c}^2}-\Omega _{\phi  c}\right)}{9 w_{\phi  c}^2 \Omega _{\phi  c}^2+6 \Omega _{\phi  c}+1}
<
\lambda_3
<
\frac{2 \left(-3 w_{\phi  c}^2 \Omega _{\phi  c}^2+\sqrt{\Omega _{\phi  c}^2-w_{\phi  c}^2 \Omega _{\phi  c}^2}-\Omega _{\phi  c}\right)}{9 w_{\phi  c}^2 \Omega _{\phi  c}^2+6 \Omega _{\phi  c}+1}\,.
\ee
For $w_{\phi c} = -0.99$, 
the above condition becomes $-0.45 < \lambda_3 < -0.41$ and $-0.51 < \lambda_3 < -0.47$ when $\Omega_{\phi c} = 0.65$ and $\Omega_{\phi c} = 0.95$, respectively.
To ensure that the scaling points are stable,
we choose $\lambda_3$ in the ranges $\lambda_3 < \lambda_{3 a}$ or $\lambda_3 > \lambda_{3 b}$.
For $w_{\phi c} = -0.99$, 
we have $(\lambda_{3 a}, \lambda_{3 b}) = (-0.57, -0.01)$ and $(-0.57, -0.13)$ when $\Omega_{\phi c} = 0.65$ and $\Omega_{\phi c} = 0.95$.
Hence, for $\lambda_3 < \lambda_{3 a}$ or $\lambda_3 > \lambda_{3 b}$, $\Omega_{\phi c}^\fy$ is real.
In the case where $w_{\phi c} \gtrsim -1$ and $\Omega_{\phi c} > 0.65$,
Eq.~(\ref{Om:y0}) gives $\Omega_{\phi c}^\fy \lesssim 10^{-3}$.
According to  the numerical values of $\lambda_{3 a}$ and $\lambda_{3 b}$,
we restrict $\lambda_3$ within the ranges $\lambda_3 \leq -2/3$ and $0 < \lambda_3 \leq 1$ in the following analysis,
where the upper bound $\lambda_{3} \leq 1$ is imposed to avoid divergence of $x'$ and $\Omega_\dm'$ which can occur when $\lambda_3$ satisfies Eq.~(\ref{l3sing}).

\begin{figure}
\includegraphics[height=0.4\textwidth, width=0.9\textwidth,angle=0]{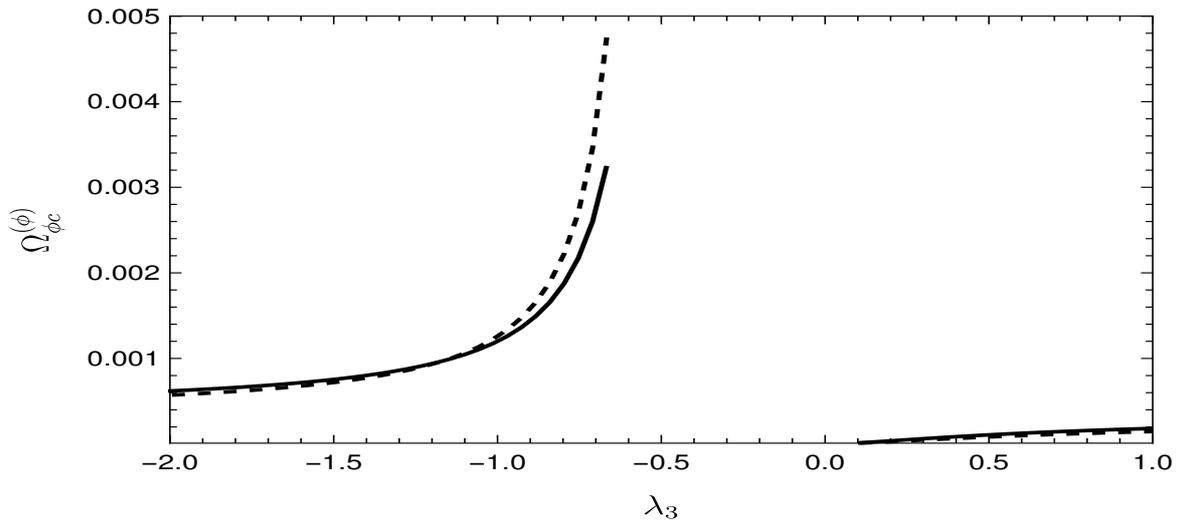}
\caption{\label{fig:2}
Plots of $\Omega_{\phi c}^\fy$ as a function of $\lambda_3$.
The solid line shows the case $\Omega_{\phi c} = 0.65$,
while the dashed line shows the case $\Omega_{\phi c} = 0.95$.
In the plots, $w_{\phi c} = -0.99$ , $\lambda_2 = 1$ and $\lambda_3$ lies within the range $\lambda_3 \leq -2/3$ and $0 < \lambda_3 \leq 1$.
The plots are not sensitive to $\lambda_2$.
}
\end{figure}

The quantity $\Omega_{\phi c}^\fy$ is the value of $\Omega_\phi$ at the $\phi$MDE point.
We plot this quantity as a function of $\lambda_3$ in Fig.~\ref{fig:2}.
We note that $\lambda_1$ in Eqs.~(\ref{l1scaling}) and (\ref{l1:y0}) can be matched only for suitable conditions for $\lambda_3$.
For example, we obtain the same expression for $\Omega_{\phi c}^\fy$ when we solve for it from the equations which are constructed by matching $\lambda_{1+}$ from Eq.~(\ref{l1scaling}) with either $\lambda_{1+}^\fy$ or $\lambda_{1-}^\fy$ from Eq.~(\ref{l1:y0}).
However, if we compute the numerical value of $\Omega_{\phi c}^\fy$ from Eq.~(\ref{Om:y0}) for given values of $\Omega_{\phi c}, w_{\phi c}, \lm_2$ and $\lm_3$, and insert the result back into Eq.~(\ref{l1:y0}),
the numerical  value of $\lambda_{1+}$ will be equal to $\lambda_{1-}^\fy$ when $\lambda_3 \leq - 2/3$ while it will be equal to $\lambda_{1+}^\fy$ when $\lambda_3 > 0$.
Moreover, $|x_{c-}^\fy| < 1$ and $|x_{c+}^\fy| > 1$ for the former case while
$|x_{c-}^\fy| > 1$ and $|x_{c+}^\fy| < 1$ for the latter case. 
The case where $|x_{c}^\fy| > 1$ is not physically relevant case.
We summarize the matching of $\lambda_1$ and $\lambda_1^{\fy}$ and the conditions on $\lambda_3$ in Tab.~\ref{tab:1}.
\begin{table}[t]
\begin{center}\begin{tabular}{|c|c|c|c|c|c|}
    \hline 
Matching Cases
 & Scaling = $\phi$MDE 
& $\lambda_3$ & $x_c^\fy$  & $\lambda_1$ & $\lambda$ 
\tabularnewline
\hline 
I &$\lambda_{1+} = \lambda_{1-}^\fy$ & $\lambda_3 \leq -2/3$ & $|x_{c-}^\fy| < 1$ and $|x_{c+}^\fy| > 1$ &$<0$ & $<0$ \\
II &$\lambda_{1+} = \lambda_{1+}^\fy$ & $\lambda_3 > 0$ & $|x_{c-}^\fy| > 1$ and $|x_{c+}^\fy| < 1$ &$>0$ & $<0$ \\
III &$\lambda_{1-} = \lambda_{1+}^\fy$ & $\lambda_3 \leq -2/3$ & $|x_{c-}^\fy| > 1$ and $|x_{c+}^\fy| < 1$ &$>0$ & $>0$ \\
IV &$\lambda_{1-} = \lambda_{1-}^\fy$ & $\lambda_3 > 0$ & $|x_{c-}^\fy| < 1$ and $|x_{c+}^\fy| > 1$ &$<0$ & $>0$ \\
\hline
\end{tabular}\end{center}
\caption{\label{tab:1}
Matching of $\lambda_1$ from Eqs.~(\ref{l1scaling}) and (\ref{l1:y0}) and the 
required conditions on $\lambda_3$.
The fourth column shows the magnitude of $x_{c}^{\fy}$. 
The fifth and the sixth columns present the signs of $\lambda_1$ and $\lambda$ computed from Eqs.~(\ref{l1scaling}) and (\ref{lmscaling}).
The main conclusions from the table do not change if $|\lambda_2| \sim {\cal O}(1)$, $w_{\phi c} \gtrsim -1$ and $\Omega_{\phi c} > 0.65$.
}
\end{table}
We now investigate the eigenvalues of the $\phi $MDE points based on the choices of parameters in Tab.~\ref{tab:1}.
The first eigenvalues of all cases are simple and are shown in Tab.~\ref{tab:2}. 
From the table, we see that the eigenvalues could be negative depending on the sign of $\lambda$.
Nevertheless, the terms $\lambda$ are multiplied by $\sqrt{\Omega_{\phi c}^\fy}$ which is in order of $10^{-2}$, 
so that these terms have no sufficient contribution to make the eigenvalues negative.
\begin{table}[t]
\begin{center}\begin{tabular}{|c|c|c|c|c|}
    \hline 
First eigenvalue &
Cases I and II & Cases III and IV
\tabularnewline
\hline 
$\mu_1$ & $ \lambda  \sqrt{6 \Omega _{\phi  c}^\fy}+3 (\Omega _{\phi  c}^\fy+1)$ & $- \lambda  \sqrt{6 \Omega _{\phi  c}^\fy}+3 (\Omega _{\phi  c}^\fy+1)$ \\
\hline
\end{tabular}\end{center}
\caption{\label{tab:2}
The first eigenvalues for all possible matching cases. 
}
\end{table}
For these $\phi $MDE points,
the polynomial for the eigenvalues is complicated.
Fortunately, the first eigenvalue takes the simple form,
so we can reduce the order of the polynomial by dividing the polynomial with $(\mu_1 - \mu)$.
The resulting polynomial is second order in $\mu$,
which can be written in the form
\be
\label{polymu}
\mu^2 + a_1 \mu + a_2 = 0\,,
\ee
where $a_1$ and $a_2$ are complicated functions of the parameters and $\Omega_{\phi c}^\fy$.
Since $\Omega_{\phi c}^\fy \lesssim 10^{-3}$,
we expand $a_1$ and $a_2$ around $\Omega_{\phi c}^\fy = 0$ up to 
$\Omega_{\phi c}^\fy$ as shown in Eqs.~(\ref{c1a1}) - (\ref{c3a2}).

\noindent
{\bf cases I and II:}
\ba
\label{c1a1}
a_1 &=&
\frac{3}{2}
-\lambda_1 \sqrt{6 \Omega_{\phi c}^\fy}
+\(-24 \lambda_3 + \frac{6}{\lambda_3} - \frac{3}{2}\)\Omega_{\phi c}^\fy + \dots\,,\\
\label{c1a2}
a_2 &=&
-\frac{9}{2}
+\frac{3 \sqrt{\frac{3}{2}} [\lambda_1 \lambda_3 \left(\lambda_3 - 5\right) + 3 \lambda_2 \left(\lambda_3 + 1\right)]}{\lambda_3 \left(\lambda_3 + 1\right)} \sqrt{\Omega_{\phi c}^\fy}
\nonumber\\
&&
 + \frac{3 \big[\lambda_1^2 \left(-2 \lambda_3^2 + 5 \lambda_3 + 1\right) - \lambda_1\lambda_2 a_{2b} - 2 \left(\lambda_2^2 \left(\lambda_3 + 1\right) - 3 \lambda_3 a_{2c}\right)\big]}{\lambda_3^2 \left(\lambda_3 + 1\right)}\Omega_{\phi c}^\fy + \dots\,,
\ea
\noindent
{\bf cases III and IV:}
\ba
\label{c3a1}
a_1 &=&
\frac{3}{2}
+ \lambda_1 \sqrt{6 \Omega_{\phi c}^\fy}
+ \(-24 \lambda_3 + \frac{6}{\lambda_3} - \frac{3}{2}\) \Omega_{\phi c}^\fy + \dots\,,
\\
\label{c3a2}
a_2 &=&
-\frac{9}{2}
-\frac{3 \sqrt{\frac{3}{2}} [\lambda_1 \left(\lambda_3 - 5\right) \lambda_3 + 3 \lambda_2 \left(\lambda_3 + 1\right)]}{\lambda_3 \left(\lambda_3 + 1\right)} \sqrt{\Omega_{\phi c}^\fy}
\nonumber\\
&&
-\frac{3 \big[\lambda_1^2 \left(2 \lambda_3^2 - 5 \lambda_3 - 1\right) + \lambda_1 \lambda_2 a_{2b} + 2 \left(\lambda_2^2 \left(\lambda_3 + 1\right) - 3 \lambda_3 a_{2c}\right)\big]}{\lambda_3^2 \left(\lambda_3 + 1\right)} \Omega_{\phi c}^\fy + \dots\,,
\ea
where $a_{2b} = 2 \lambda_3^2 - 3 \lambda_3 + 1$ and $a_{2c} = 2 \lambda_3^3 - 2 \lambda_3^2 + 3 \lambda_3 + 1$.
The solutions of Eq.~ (\ref{polymu}) are 
\begin{equation}
\mu_{\pm} = \frac{-a_1 \pm \sqrt{a_1^2 - 4 a_2}}{2} \,. 
\end{equation}
From these solutions we see that the real part of at least one solution is negative if $a_1>0$.
If $a_1<0$, the real part of one solution is negative when $a_2<0$. 
According to Eqs.~(\ref{c1a1}) and (\ref{c3a1}) and the sign of $\lambda_1$ 
in Tab.~\ref{tab:1}, 
the main contributions to $a_1$ for the cases I and III are positive.
As a result, the real part of at least one eigenvalue for each case is negative.
For the cases II and IV, it follows from Eqs. (\ref{c1a2}) and (\ref{c3a2}) 
together with the sign of $\lambda_1$ and the range of $\lambda_3$ in Tab.~\ref{tab:1} that the main contributions to $a_2$ can be negative.
However, to ensure that $a_2$ is negative,
we suppose that $|\lm_2| < |\lm_1|$ and impose the additional condition $\lm_3 \leq 1$ which is required to avoid divergence of $x'$ and $\Omega_\dm'$.
This suggests that the real part of one eigenvalue for each case is negative.
From the above discussion,
we conclude that the $\phi$MDE point can be saddle for $\lambda_3$ given in the table, $\lm_3 \leq 1$, $|\lambda_2| \sim {\cal O}(1)$ and for $w_{\phi c}$, $\Omega_{\phi c}$ satisfying the observational bound, e.g., $w_{\phi c} = -0.99$ and $\Omega_{\phi c} > 0.65$.

\subsection{Evolution from the $\phi$MDE point to scaling point}

\begin{figure}
\includegraphics[height=0.4\textwidth, width=0.9\textwidth,angle=0]{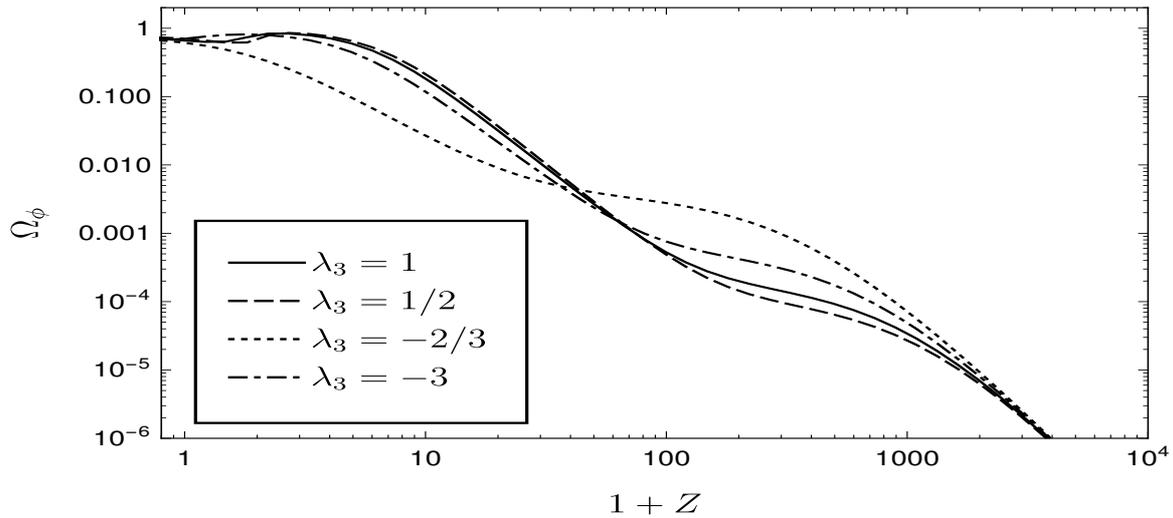}
\caption{\label{fig:3}
Evolutions of $\Omega_{\phi}$ for various values of $\lambda_3$.
In the plots, $1 + Z = 1/a$.
}
\end{figure}

\begin{figure}
\includegraphics[height=0.5\textwidth, width=0.49\textwidth,angle=0]{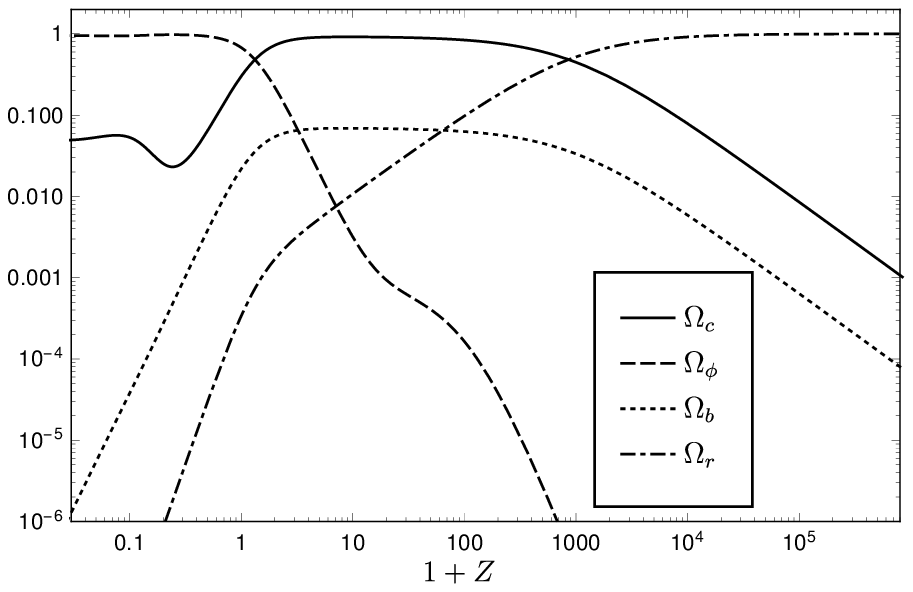}
\includegraphics[height=0.5\textwidth, width=0.49\textwidth,angle=0]{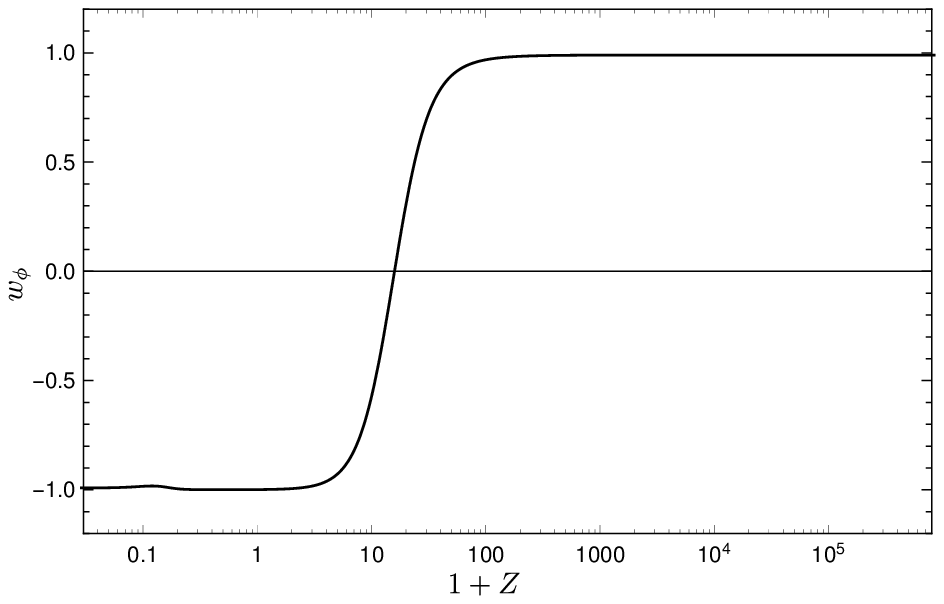}
\caption{\label{fig:4}
The left panel shows the evolutions of $\Omega_r, \Omega_b, \Omega_c$ and $\Omega_\phi$,
while the right panel shows the evolution of $w_\phi$.
The $\phi$MDE takes place around $1 + Z \sim 20$.
}
\end{figure}

We now numerically study the evolution of the background universe through the fixed points discussed in the previous sections.
The evolution equations used in the numerical integration are obtained by substituting Eq.~(\ref{CoefC}) into Eqs.~(\ref{xp}) -- (\ref{omp}).
To illustrate some results in the previous sections,
the evolutions of $\Omega_\phi$ for various values of $\lambda_3$ are plotted in Fig.~\ref{fig:3}.
In the figure,
we set $\lambda_2 = 1$, $\Omega_b = 0$ and specify $\lambda$ and $\lambda_1$ by setting $\Omega_{\phi c} = 0.7$ and $w_{\phi c} = -0.99$.
From the figure,
we see that the fixed point $\Omega_\phi = \Omega_{\phi c} = 0.7$ can be reached at late-time.
From the numerical investigation,
the whole evolution of $\Omega_\phi$ weakly depends on $\lambda_2$,
and the late-time evolution is  robust under the change of initial conditions.
We next add the contribution from the energy density of baryon into the numerical integration by setting $\Omega_b \simeq 0.022$ at present.
The evolutions of $\Omega_r$, $\Omega_\dm$ and $\Omega_\phi$ for $\lambda_3 = -3/2$ are plotted in Fig.~\ref{fig:4}.
In these plots,
we set $\lambda_2 = 1$.
The parameters $\lm$ and $\lm_1$ are specified by setting $\Omega_{\phi c} = 0.95$ and $w_{\phi c} = -0.99$.
We set $\Omega_{\phi c}$ to be larger than the observational bound for the present value of $\Omega_\phi$ because this scaling point can be reached in the future when $\Omega_b \sim 0$.
From the figure we see that  the universe evolves from the radiation domination to $\phi$MDE point and then evolves towards the scaling point at late time with $\Omega_\phi \to 0.95$ and $\Omega_b \to 0$.
This pattern of the evolution is achieved for wide ranges of $\lambda_2$ and initial conditions.                      
Before reaching the late-time attractor,
the cosmic evolution can pass the point $\Omega_\phi \simeq 0.68, \Omega_\dm \simeq 0.3$ and $\Omega_b \simeq 0.022$ at present as required by observations.

\section{Growth of density perturbations}
\label{sec4}

In this section, we consider the growth of density perturbations of matter on small scales. 
To compute the evolution equations for the perturbations,
we use the metric perturbation in the Newtonian gauge written in the form
\be
ds^2 =  - \(1+ 2\Psi\) dt^2 + a^2\(1 - 2\Psi\)\delta_{ij} dx^i dx^j\,,
\ee
where $\Psi$ is the metric perturbation and the anisotropic perturbations are omitted.
The field $\phi$ is decomposed into the background and perturbed parts as
\be
\phi \to \phi + \de \phi\,,
\ee
where $\phi$ on the right-hand side of the arrow denotes the homogeneous background part of the field while $\de\phi$ denotes the perturbed part.
Applying these decompositions to Eq.~(\ref{gkg-qg}),
we obtain the evolution equation for the field perturbations as
\be
\delta\ddot\phi + 3 H \left(\delta\dot\phi - 2 \Psi \dot{\phi}\right)
+ \left(V_{, \phi  \phi }+\(\frac{k}{a}\)^2\right)\delta\phi
-2 \ddot{\phi } \Psi - 4 \dot{\phi } \dot{\Psi } =  \delta Q\,,
\label{kg:pert}
\ee
where the perturbations in the coupling $\delta Q$ is given by
\ba
\delta Q &=& \delta Q_S 
+ \(- \tQ \dot{\phi} \tCXp + \ddot{\phi}(2 X \tCXpX + \tCXp) + 2 X \tCXpp - \tCpp\) \delta\phi
\nonumber\\
&& 
+ \Big[- \tQ \Theta + \dot{\phi} \left(2 X \tCXpX + 2 \tCXp - \tCpX \right) + \ddot{\phi} \left(2 X \tCXXX + 3 \tCXX\right)\Big] \delta\dot\phi
\nonumber\\
&&
+ \Theta \delta\ddot\phi - 6 H \dot{\phi} \left(X \tCXX + \tCX\right)\Psi 
- \tCX(2 \ddot{\phi} \Psi + 4\dot{\phi} \dot{\Psi})\nonumber\\
&& - 2 X \left(\Big(\dot{\phi} \tCXX \frac{\dot{\rho}_\dm}{\rho_\dm} + 2 X \tCXpX + 2 \tCXp - \tCpX + 2 X \tCXXX \ddot{\phi} + \tCXX \ddot{\phi}\Big)\Psi + \dot{\phi} \tCXX \dot{\Psi} \right) \,.
\ea
Here,  $\delta Q_S$ contains the dominant contribution to $\delta Q$ on small scales which its expression is given by 
\be
\delta Q_S = \tCX \(\frac{k}{a}\)^2\delta\phi
+ \[- 2 \tQ X \tCX + 2 X \tCXp - \tCp + \Theta \ddot{\phi}\]\delta_\dm
+ \dot{\phi} \tCX \dot\delta_\dm\,,
\label{dqsmall}
\ee
where $k$ is the comoving wave number of the perturbation modes.
The density contrast $\delta_\dm \equiv \delta\rho_\dm / \rho_\dm$, where $\delta\rho_\dm$ and $\rho_\dm$ are the perturbations in energy density and background energy density of CDM.
The term $\delta Q_S$ is obtained from the fact that on small scales,
$|\delta\ddot\phi|$ and $|H\delta\dot\phi|$ are much smaller than $|k^2 \delta \phi / a^2|$
and $|H \delta_\dm|$ as well as $|\dot\delta_\dm|$ are much larger than the $\Psi$ terms.
The latter approximation follows from the perturbed Einstein equation on small scales: 
\be
\(\frac{k}{a}\)^2 \Psi = - \frac{3}{2} H^2 \(\Omega _\dm \delta_\dm + \Omega_b \delta_b\)\,,
\ee
where $\de_b \equiv \de\rho_b /\rho_b$ is the density contrast of baryon.
In the above equation,
the small contributions from the perturbations in the energy density of radiation and dark energy are neglected.
On small scales,
Eq.~(\ref{kg:pert}) becomes
\be
\(\frac{k}{a}\)^2 \delta\phi=  \delta Q_S\,.
\ee
Combining the above equation with Eq.~(\ref{dqsmall}),
we get
\be
\(\frac{k}{a}\)^2 \delta\phi=  \rho_\dm\tDQ\,,
\ee
where the effective coupling term in the perturbed universe is
\be
\tDQ = \frac{\left( 2 X \tQ \tCX 
-2 X \tCXp + \tCp - \Theta \ddot{\phi }\right) \delta_\dm
-\dot{\phi } \tCX \dot{\delta _\dm}}{\tCX \rho_\dm-1}\,.
\label{tdq} 
\ee
We see that there is the term that is proportional to	 $\dot\de_\dm$ in the effective coupling term,
this term vanishes when the transformation coefficient $C$ does not depend on $X$. 
The evolution equations for the perturbations in energy density and velocity $v_\dm$ of CDM are computed from Eq.~(\ref{grdmatt}).
The resulting equations are given by
\ba
{} && \dot{\delta _\dm} - 3 \dot{\Psi } - \(\frac{k^2}{a}\) v_\dm
=
 \dot\phi \tQ \delta _\dm
- \dot{\phi} \frac{\delta Q}{\rho_\dm} 
- \tQ \delta\dot\phi \,,
\label{eqdm}\\.
{} && \dot{v}_\dm + \(H - \dot\phi\tQ\) v_\dm + \frac 1a \Psi
=
\frac{\tQ}{a} \delta\phi\,.
\label{eqvm}
\ea
Since we concentrate on small scales perturbations,
we replace $\delta Q / \rho_\dm$ in Eq.~(\ref{eqdm}) by $\tDQ$ from Eq.~(\ref{tdq}) and keep only the dominant terms on small scales.
The resulting equation is
\be
\dot{\delta _\dm} - \(\frac{k^2}{a}\) \frac{\tCX \rho_\dm-1}{\tCX \rho_\dm - 2 X\tCX - 1} v_\dm
=
 \dot\phi \frac{\Theta V_{, \phi } + 3 H \Theta \dot{\phi } -2 X \tCXp + \tCp}{\Theta\rho_\dm -2 X \tCX - 1} \delta _\dm
+ \frac{2 \dot\phi X \tCXp - \dot\phi \tCp + \Theta \dot\phi\ddot{\phi }
}{\tCX \rho_\dm - 2 X\tCX - 1}\delta_\dm
\equiv \Delta \de_\dm\,.
\label{eqdm-sim}
\ee
It can be checked that if the coefficient of the conformal transformation solely depends on the field $\phi$,
$\Delta$ vanishes.
Hence, $\Delta$ describes effects of the coupling due to general conformal transformation on the growth of matter perturbations.
To discuss the influences of $\Delta$ on the matter perturbations,
we suppose that $|\lm_2| \sim {\cal O}(1)$, $\lm_3$ lies with in the ranges shown in Tab.~\ref{tab:1} and the evolution of $z$ does not much deviate from the fixed point $z = \lm_3 / x^2$.
Based on this assumption,
we have $H^2 \tCX = z / (2 + 6 z x^2) \sim {\cal O}(1/ (6 x^2))$ and $2 h^2 X \tCXX \sim- 4 / (3 x^2) $,
so that $\Theta$ is negative and $|\rho_\dm\Theta| \sim |\rho_\dm \tCX| \sim {\cal O}(\Omega_\dm / x^2) \gg 1$. 
For $\tCp$, we have $\tCp \sim \lm_1 / (2 + 6 z x^2) \sim \lm / (6\lm_3)$,
while we have $|X \tCXp |\ll 1$.
From Eq.~(\ref{l1scaling}), we can check that the second term on the right hand side is the dominant term because we always set $w_{\phi c} = -0.99$.
To simplify our analysis,
we suppose that $\Omega_{\phi c} > 0.65$ and impose the additional condition for the possitive $\lm_3$ to be $1/2 \leq \lm_3 \leq 1$.
Hence, Eq.~(\ref{l1scaling}) implies that $\lm_{1+}$ has the same sign as $\lm_3$ 
while $\lm_{1-}$ and $\lm_3$ have opposite signs. 
As a result,
$\tCp$ is positive when $\lm_1 = \lm_{1+}$ and becomes negative when $\lm_1 = \lm_{1-}$.
From our numerical investigation,
the sign of $\dot\phi$ is preserved through the cosmic evolution.
It follows from Eqs.~(\ref{l1scaling}) and (\ref{xyscaling}) that $\dot\phi$ is positive for $\lm_1 = \lm_1+$ and negative for $\lm_1 = \lm_{1-}$.
Hence, the term $\dot\phi \tCp$, which gives the  dominant contribution to the numerators of $\Delta$,
is positive.
For the denominators of $\Delta$,
the dominant contributions come from $\Theta\rho_\dm$ and $\tCX \rho_\dm $ which are negative and positive, respectively.
According to the above analysis,
the dominant contribution to $\Delta$ is negative and therefore this term suppresses the growth of $\de_\dm$.

To perform the numerical investigation,
we compute the evolution equation for the CDM perturbation $\delta_\dm$ by differentiating Eq.~(\ref{eqdm-sim}) with respect to time.
The time derivative of $v_\dm$ in the resulting equation is eliminated using Eq.~(\ref{eqvm}).
Finally, the remaining $v_\dm$ terms can be eliminated using Eq.~(\ref{eqdm-sim}),
and we get
\ba
\delta_\dm'' + C_1 \delta_\dm' 
-\frac{3}{2} \left( G_{cc} \Omega_\dm \delta_\dm 
+ G_{cb} \Omega_b \delta_b\)
= 0\,,
\label{eqddm}
\ea
where $C_1$, $G_{cc}$ and $G_{cb}$ are the functions of $x, y, z, \Omega_\dm, \Omega_b$ and parameters of the model.
The expressions for these coefficients, especially $G_{cc}$, are lengthy,
so that their explicit forms are not shown here.
Since $\Omega_\dm > \Omega_b$,
the contribution to the evolution of  $\de_\dm$ from $G_{cc}$ is larger than that from $G_{cb}$.
Hence, the effective gravitational coupling relevant to CDM perturbations on small scales is dominated by $G_{cc}$.
\begin{figure}
\includegraphics[height=0.4\textwidth, width=0.9\textwidth,angle=0]{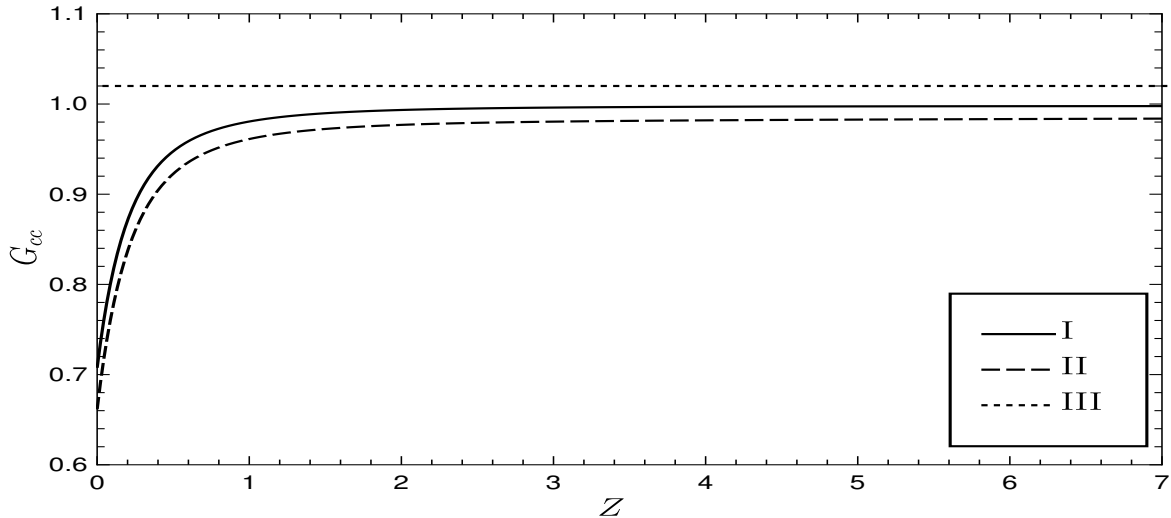}
\caption{\label{fig:5}
Plots of $G_{cc}$ as a function of $Z$.
The lines I and II represent the cases where $(\lm_3, \Omega_{\phi c}, w_{\phi c}) = (1/2, 0.96, -0.99)$ and $(-3/2, 0.99, -0.99)$, respectively.
For these lines, $\lambda_2 = 1$.
The line III represents the case of usual conformal coupling $(z = 0)$ with $(\lambda,\lambda_1) = (-1/10,-2/10)$.
In all plots, $\Omega_b \simeq 0.022, \Omega_\dm \simeq 0.3$ and $\Omega_\phi \simeq 0.68$ at present.
}
\end{figure}
The numerical value of $G_{cc}$ is shown in Fig.~\ref{fig:5}.
We see that after the matter dominated epoch, $G_{cc}$ for the coupled model from general conformal transformation is smaller than unity, 
while $G_{cc}$ from the usual conformal coupling is larger than unity.
Since $G_{cc} =1$ for $\Lambda$CDM model,
the effective gravitational coupling is suppressed in the coupled dark energy model inspired from general conformal transformation.
This suggests the weaker growth of  CDM perturbations on small scales which can be estimated by numerically solving Eq.~(\ref{eqddm}).
In order to solve Eq.~(\ref{eqddm}),
we have to know the evolution equation for $\de_b$.
Since the energy-momentum tensor of the baryon is separately conserved,
the evolution equation for the density contrast of baryon on small scales takes the usual form as
\be
\delta_b'' + \(2 + \frac{\dot{H}}{H^2}\)  \delta_b'
- \frac 32 \(\Omega_b \delta_b + \Omega_\dm \delta_\dm \) 
 = 0\,.
\label{eqddb}
\ee
We solve Eqs.~(\ref{eqddm}) and (\ref{eqddb}) numerically based on the parameters in Fig.~\ref{fig:5}.
The evolutions of $\de_c / a$ are plotted in Fig.~\ref{fig:6}.
In the figure,
we normalize $\de_\dm / a$ such that it is unity at the present.
From the figure,
we see that $\de_\dm / a$ for the coupled dark energy model from general conformal coupling is larger than that for $\Lambda$CDM model in the early epoch.
This implies that the growth of CDM perturbations is weaker in this coupled dark energy model.
To estimate how much the $\sigma_8$ tension can be resolved in the coupled dark energy model with general conformal coupling,
we have to perform a likelihood analysis which we leave for a future work.

 \begin{figure}
\includegraphics[height=0.4\textwidth, width=0.9\textwidth,angle=0]{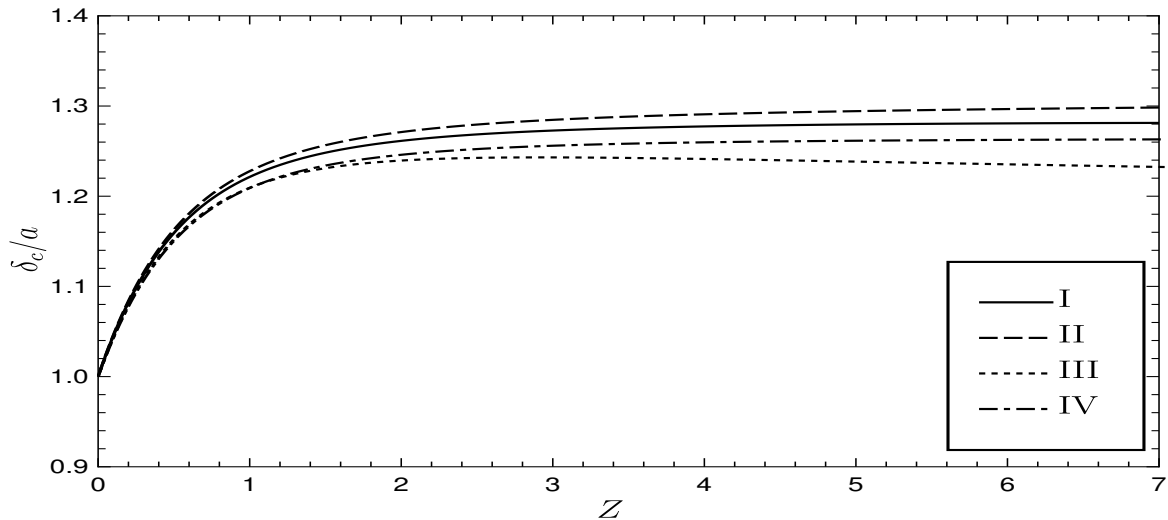}
\caption{\label{fig:6}
Evolutions of $\delta_c/a$. 
The choices of parameters for the lines I, II and III are the same as those for the lines in Fig.~\ref{fig:5}.
The line IV represents the case of $\Lambda$CDM with $\Omega_b \simeq 0.022, \Omega_\dm \simeq 0.3$ and $\Omega_\Lambda \simeq 0.68$ at present.
}
\end{figure}

\section{Conclusions}
\label{sec5}

In this work,
we have studied coupled dark energy model inspired from general conformal transformation in which the transformation coefficient depends on both scalar field and its kinetic term.
The effective coupling term consists of the multiplication between the derivative of the scalar field and the energy density as well as between the derivative of the scalar field and derivative of energy density of CDM which can lead to different influences on the growth of matter perturbations compared with the usual conformal coupling case.

The scaling solutions can exist in this coupled dark energy model.
The solution which corresponds to the $\phi$MDE can be a saddle point,
while the solution for the cosmic acceleration at late time can be attractor.
The background universe can evolve from the radiation dominated epoch through the $\phi$MDE towards the cosmic acceleration epoch at late time. 
This sequence of the evolution can be achieved for the cosmological parameters which satisfy the observational bounds.
The existence of the $\phi$MDE modifies the effective equation of state parameter during the matter dominated epoch,
such that the $H_0$ from the CMB analysis for this model could be larger than that for the $\Lambda$CDM model,
which potentially solves the $H_0$ tension.
However, the actual likelihood analysis is required to estimate how much the $H_0$  tension can be resolved.

The growth of the linear matter perturbations on small scales for this coupled dark energy model is weaker than that for the $\Lambda$CDM model.
This is a consequence of the reduction of the effective gravitational constant relevant to the CDM perturbations on small scales. 
The suppression of the growth of CDM perturbations on small scales suggests that the $\sigma_8$ tension could be alleviated in this model.  
However, to investigate whether  this model of coupled dark energy can actually solve the $H_0$ and $\sigma_8$ tensions,
a full likelihood analysis is needed which we leave for a future work. 

\subsection*{Acknowledgement}

WT was supported by Royal Thai Government Scholarship (Ministry of Higher Education, Science, Research and Innovation) for his Ph.D. study.
KK is supported by Fundamental Fund from National Science, Research and Innovation Fund under the grant ID, P2565B202.
The authors thank the referee for a comment on a momentum transfer.


\end{document}